# Crazing of Nanocomposites with Polymer-Tethered Nanoparticles

**Dong Meng\* and Sanat K. Kumar**
*Department of Chemical Engineering, Columbia University,
New York, NY 10027*

**Ting Ge**
*Department of Chemistry, University of North Carolina, Chapel Hill, NC 27599*

**Mark Robbins**
*Department of Physics and Astronomy, Johns Hopkins University, Baltimore, MD, 21218*

**Gary S. Grest**
*Sandia National Laboratories, Albuquerque, NM 87185*

## Abstract

The crazing behavior of polymer nanocomposites formed by blending polymer grafted nanoparticles with an entangled polymer melt is studied by molecular dynamics simulations. We focus on the three key differences in the crazing behavior of a composite relative to the pure homopolymer matrix, namely, a lower yield stress, a smaller extension ratio and a grafted chain length dependent failure stress. The yield behavior is found to be mostly controlled by the local nanoparticle-grafted polymer interfacial energy, with the grafted polymer-polymer matrix interfacial structure being of little to no relevance. Increasing the attraction between nanoparticle core and the grafted polymer inhibits void nucleation and leads to a higher yield stress. In the craze growth regime, the presence of "grafted chain" sections of $\approx 100$ monomers alters the mechanical response of composite samples, giving rise to smaller extension ratios and higher drawing stresses than for the homopolymer matrix. The dominant failure mechanism of composite samples depends strongly on the length of the grafted chains, with disentanglement being the dominant mechanism for short chains, while bond breaking is the failure mode for chain lengths $> 10 N_e$, where $N_e$ is the entanglement length.

\*Current address: *Dave C. Swalm School of Chemical Engineering, Mississippi State University, Starkville, MS 39762*

## I. Introduction

It is well known that the addition of nanoscale fillers to polymer materials can lead to markedly enhanced mechanical properties. Polymer nanocomposites (PNC) exhibit improved mechanical properties in processing, modulus, strain-to-failure, and toughness, relative to pure polymers and also to composites comprised of conventional microscale fillers.[1-5] The reinforcing mechanism of adding nanofillers has been under intense study in the past decade. It is now appreciated that changes to polymer behavior depend on numerous properties including particle concentration, particle geometry, particle size, interfacial interactions, and thermal history. While extensive work has been dedicated to studying the low-strain elastic behavior of PNC,[6] little attention has been paid to the equally important plastic flow regime that constitutes the major part of the mechanical performance before fracture. In particular, glassy polymeric materials exhibit unique mechanical failure through crazing, where undeformed polymer evolves into an intricate network of fibrils called crazes. Crazing dissipates an exceptional amount of energy and increases the fracture energy by factors of several thousand, making polymer glasses desirable for load-bearing purposes in many engineering applications. The molecular origins and dynamic development of crazes in pure polymers have received considerable attention experimentally[7, 8] and theoretically,[9-12] but attempts to extend the current understanding to nanocomposite materials have been limited.

Papakonstantopoulos et al.[13, 14] demonstrated a strong correlation between local mechanical properties and the nature of nonaffine plastic failure of nanocomposites through computer simulations. They found that the average moduli of PNC filled with NPs is larger than that of the unfilled polymer. Experimentally Lee et al.[15] reported changes in the distribution of nanoparticles (NPs) in glassy PNCs during craze development. They found alignment of surface-treated NPs along the pre-craze, expulsion of NPs from craze fibrils of the pre-craze, and NP entrapment among craze fibrils in the mature craze. A further study[16] also concluded that NPs at the craze-bulk interface can serve as separators between polymer chains because of the increased mobility of polymer segments at an enthalpically neutral NP surface. Therefore, the probability for chains to form pile-up entangled strands at the craze-bulk interface decreases as NP loading increases. Consequently, less cross-tie fibrils form in the craze, and greater strain is required to exceed the critical fracture energy for crack propagation. Lin et al.[17] investigated the crazing of glassy polystyrene (PS) mixed with polystyrene-grafted multiwalled carbon nanotubes (PS-MWCNT). They also found a high local concentration of MWCNTs at craze boundaries. The pile up is thought to result from the incapability of the softened chains to pull the rigid CNTs into the nanoplastic flow. As crazes widen, the polymer chains drawn into crazes are being filtered through the pileup of MWCNTs at craze boundaries, causing a significant increase of chain friction during micro deformation. As the result, craze widening becomes progressively difficult, and ultimately enhances delocalization of the plastic flow of crazing.

Toepperwein and de Pablo[18] investigated the early stages of crazing in nanocomposites of linear polymers and nanorods using computer simulations. They found that voids form preferentially in regions of low local elastic modulus, and that the addition of attractive NPs induces earlier void formation due to a more mechanically heterogeneous environment. In the developing crazes, they find that larger particles resist incorporation into developing voids, consistent with the trapped regions of additive found experimentally by Lee et al.[15, 16] Riggleman et al.[19] showed that bare NPs can serve as entanglement attractors, particularly at large deformations, altering the topological constraint network that arises in the composite material. Gersappe[20] showed that, at the same loading of nanofillers, smaller fillers increase the area between fillers and polymer matrix and thus improve the toughness of the composite above the glass transition $T_g$. These simulations also showed that the increase in the attraction strength between nanofillers and polymer matrix enhances the composite toughness above $T_g$. However, Gersappe found little improvement in the composite toughness below $T_g$.

In this study, we focus on PNCs of long entangled polymer chains mixed with polymer-grafted (or brush) NPs. Grafting polymers to NPs is an effective strategy for improving the dispersion of NPs in polymer nanocomposites. In contrast to surfactant coated NPs, polymer chains grafted onto NPs are typically of the same chemistry as the surrounding matrix. Entropic effects, which are controlled by the grafting density and brush-matrix chain length ratio, modify the effective interaction between polymer grafted NPs,[21, 22] and can be used for controlling their state of dispersion in a polymer matrix. Studies have shown that adjusting grafting density and the chain length ratio of the graft to the matrix chains allowed for control over the self-assembly of polymer grafted NPs, e.g., structures such as spheres, strings and sheets can be formed.[23] Given the different levels of brush-brush and brush-matrix interactions as the result of varying the spatial arrangements of NPs, the composites are expected to exhibit distinct mechanical properties, opening up new avenues for producing hybrid materials with "designer properties".

We study the crazing behavior of composites comprised of polymer-grafted NPs using molecular dynamics (MD) simulations. Figure 1 shows snapshots comparing crazing for nanocomposites and pure polymer samples. We focus on understanding the role played by polymer grafted NPs in changing the crazing behavior of nanocomposites as compared to that of neat polymers, and PNCs composed of bare NPs. In particular, we consider the regime where the NPs are well mixed with the matrix, i.e., when the chain length of the end-grafted polymer is comparable to or greater than that of the matrix chains.[24-26] Specifically, we characterize the differences in the three stages of craze development: the cavity nucleation regime that determines the yield stress $\sigma_y$, the craze growth regime where material is deformed into a craze at a constant drawing stress $\sigma_d$ and extension ratio $\Lambda$, and the fracture regime where the fully formed craze grows until bonds break or chains disentangle at the maximum stress $\sigma_{max}$. We offer insights into the molecular level mechanisms that are responsible for these differences. Our study suggests a strong dependence of the mechanical properties of nanocomposites on the nature of surface-tethered polymers.

In the next section we define the model and methodology used in this study. In Sec III, we present results for the static structures prior to deformation and for the stress-strain curves. In Secs IV-VI, we present our results for cavity nucleation, growth and failure. Finally, in Sec. VII we present a brief summary of our main conclusions.

## II. Model and Methodology

The chains and the NPs are all assumed to be formed by chemically identical monomers with a mass $m$ and diameter $\sigma$. All monomers including bonded monomers interact through the truncated and shifted Lennard-Jones potential

$$U_{m,m}(r) = \begin{cases} 4\varepsilon\left[\left(\frac{\sigma}{r}\right)^{12} - \left(\frac{\sigma}{r}\right)^{6}\right] - 4\varepsilon\left[\left(\frac{\sigma}{r_c}\right)^{12} - \left(\frac{\sigma}{r_c}\right)^{6})\right] & r \leq r_c \\ 0 & r > r_c \end{cases} \quad (1)$$

where $\varepsilon$ is the well-depth. The cutoff distance, $r_c = 2^{1/6}\sigma$ or $1.5\sigma$.

Nanoparticles are smooth spheres of radius $a = \sigma_{NP}/2 = 10\sigma$ and their mass and interactions are calculated by integrating over their volume and assuming they contain monomer-like Lennard-Jones atoms with number density, $\rho\sigma^3=1.0$. The interaction energy between a pair of NPs has the form: $U_{C,C}(r) = U^A_{C,C}(r) + U^R_{C,C}(r)$. The attractive component $U^A_{C,C}(r)$ is given by

$$U^A_{C,C}(r) = -\frac{A_{CC}}{6}\left[\frac{2a^2}{r^2-4a^2} + \frac{2a^2}{r^2} + \ln\left(\frac{r^2-4a^2}{r^2}\right)\right] - U^A_{shift} \tag{2}$$

and the repulsive component $U^R_{C,C}(r)$ has the form of

$$U^R_{C,C}(r) = \frac{A_{CC}}{37800}\frac{\sigma^6}{r}\left[\frac{r^2-14ra+54a^2}{(r-2a)^7} + \frac{r^2+14ra+54a^2}{(r+2a)^7} - \frac{2(r^2-30a^2)}{r^7}\right] - U^R_{shift}. \tag{3}$$

$U_{shift}^{R,A}$ are the values of the potential at the cutoff radius $R_c$. The Hamaker constant $A_{CC} = 4\pi^2\varepsilon_{CC}\rho^2\sigma^6 = 39.478\varepsilon$ for NPs made of the same type of monomers as the polymer chains. The cutoff radius is selected to be $R_c = 20.45\sigma$ so that NP/NP interactions are purely repulsive.

The interaction between a NP and a chain monomer is given by:

$$U_{C,m}(r) = \frac{2a^3\sigma^3 A_{CS}}{9(a^2-r^2)^3}\left[1 - \frac{(5a^6+45a^4r^2+63a^2r^4+15r^6)\sigma^6}{15(a-r)^6(a+r)^6}\right] - U_{shift}, \tag{4}$$

where $U_{shift}$ is the value of the potential at $r = R_C$. Here we set $A_{CS} = 80\varepsilon$. Unless stated otherwise, the cut-off radius for the polymer/NP interaction is $R_c = 11\sigma$, in which case the NP-polymer interactions are purely repulsive. In a few cases, $R_c$ is increased to $15\sigma$ to study the effects of attractive interactions between the NP and polymer chains.

To represent the catenation of chains we adopt two types of bonded interactions depending on the stages in the simulation. The finitely extensible nonlinear elastic (FENE) potential is used during equilibration, with spring constant $k = 30\varepsilon/\sigma^2$ and a maximum bond extent $R_0 = 1.5\sigma$. To allow covalent bonds to break, "FENE" bonds are replaced by "quartic" bonds

$$U_{quartic}(r) = K(r-R_0)^3(r-B) + U_0 \tag{5}$$

in the quenching and active deformation stages. $R_0 = 1.5\sigma$ is the maximum bond length beyond which bonds become permanently broken. Three other constants: $K = 2351\varepsilon/k_B$, $B = 0.7425\sigma$ and $U_0 = 94.745\varepsilon$, ensure that the average bond length $l_b = 0.97\sigma$ matches the FENE bond value.

The maximum force before a covalent chain bond breaks is set to be 100 times higher than that required to break a Lennard-Jones bond, as in previous simulations[10, 27-29] and supported by experiment.[30, 31] In addition, a bond bending potential $U_{bend}(\theta) = k'\cos\theta$ is also applied to increase the degree of entanglement in the system, where the bending energy constant $k' = 0.75\varepsilon$ and θ is the angle between two neighboring bonds. This set of parameters corresponds to an entanglement length $N_e \sim 40$, and a persistence length $l_p = 2.10\sigma$.[32]

Initially, 50 NPs of diameter $\sigma_{NP} = 20\sigma$ are randomly placed inside a simulation cell under the constraint that no overlap is allowed between any two NPs. Then 125 polymer chains are randomly grafted to the surface of each NP, corresponding to a grafting density of $\Sigma = 0.1$ chains/$\sigma^2$. A total of $n$ free, matrix polymer chains are created inside the box using the method of Auhl et al.[33] In all cases no overlap was allowed between the monomers and the NPs; monomer-monomer overlap was allowed. The overall (grafted and matrix) monomer density was chosen to be $\rho_m = 0.85\sigma^{-3}$ with the additional restriction that no monomers overlap the NPs. In this study, the grafted chain length assumed values of $N_g = 200$, 400 and 600 in a series of simulations while the degree of polymerization of the matrix chains $N = 400 \sim 10N_e$ so that the matrix chains are well-entangled. The initial volume fraction $\eta_{NP} \equiv \frac{\pi\sigma_{NP}^3 N_{NP}}{6V}$ of NPs is ~5%, resulting a simulation box period of $L\sim170\sigma$ and a total of ~4 million "beads" in each simulation. Overlapping monomers in the initial states are pushed off each other using a soft potential until they are far enough apart for the LJ interaction to be switched on.

In the equilibration stage, double bridging MC moves are first performed for $10^5$ τ to relax global chain conformations at constant pressure with $T = 1.0\varepsilon/k_B$ and $P = 5.0\epsilon/\sigma^3$ with the cut-off distance $R_c = 2^{1/6}\sigma$.[33] The equilibration process continues without the double bridging moves for another $10^5$ τ at constant volume with $R_c = 1.5\sigma$, to further relax local structures. The final pressure of the equilibrated samples $P \sim 1.9\varepsilon/\sigma^3$. The samples are then quenched at constant volume at a rate of $0.0004\varepsilon/k_B\tau$ to the temperature where the hydrostatic pressure becomes zero, followed by further cooling at the same rate to the final temperature $T = 0.2\ \varepsilon/k_B$ at $P \sim 0$. A Langevin thermostat with a damping constant $\Gamma = 1.0\tau^{-1}$ was employed during equilibration and cooling to maintain the temperature. The glass temperature $T_g$ of all the samples is 0.35-0.37$\varepsilon/k_B$ as determined from the break in slope of the density vs. temperature plots shown in

Figure 2. Therefore, the working temperature T=0.2 $\varepsilon/k_B$ for mechanical tests is well below $T_g$ for all the samples.

In this study, samples are subject to active deformation following the quenching without any aging. Since strong triaxial tensile stresses are in general required to induce cavitation and crazing of polymer glasses,[7] the box dimensions $L_x$ and $L_y$ are kept constant during mechanical deformation while $L_z$ is stretched, accompanied by proportional changes in the corresponding coordinates of all particles in the box. Samples are deformed from the isotropic state with an acceleration of $\frac{d^2 L_z}{dt^2} = 5 \times 10^{-5} \sigma/\tau^2$ until the deformation rate reaches $dL_z/dt = 0.02\sigma/\tau$ and then the deformation rate is held constant afterward. The normal stresses $\sigma_{\alpha\alpha}$ are recorded during deformation as a function of the stretching factor $\equiv L_z/L_z^0$, where $L_z^0$ is the simulation box size at the isotropic state and $\alpha = x, y, z$. The stresses depend weakly (logarithmically) on the deformation rate.[11]

### III. Static Structures and Stress-Strain Curves

Before discussing the mechanical response, it is helpful to examine the static structure of the samples in their isotropic, glassy states. Figure 3 shows radial densities of polymer segments around a NP for samples with $N_g = 200, 400$ and $600$. The multiple peaks near the NP surface suggest layering of primarily grafted polymer segments. While the overall polymer density profiles are similar, the degree of brush-brush interpenetration increases with increasing grafting chain length $N_g$. Since the density of free polymers decreases as $N_g$ increases, the number of brush-matrix contacts decreases with increasing $N_g$. For $N_g = 600$, free polymers make up only ~10% of the total monomers in the sample, and the brush-brush interactions become dominant. Under the crude approximation that the entanglement density is proportional to the product of the segmental densities, the amount of entanglements formed between brushes increases with increasing grafted chain length at the cost of brush-matrix entanglements at fixed NP core volume fraction.[34]

The stress along the stretching direction $\sigma_{zz}$ is shown as a function of the stretching factor $L_z/L_z^0$ in Figure 4. As discussed in previous work[11] the stress-strain curves of crazing glassy polymers consist of four regimes: cavity nucleation in the range $L_z/L_z^0 \leq 1.3$, craze growth by pulling fibers from uncrazed material at fixed drawing stress σ_d and extension ratio Λ ($L_z/L_z^0$ up to Λ), stretching of the craze under increasing stress and finally craze failure starting at the peak stress. The system initially responds homogeneously with a stress that rises rapidly with strain. The stress begins to drop when cavities nucleate at σ_yield, allowing the surrounding material to relax back towards the equilibrium density. The cavities localize on a plane and craze growth initiates. The top panel of **Figure 1** shows a snapshot during craze growth. Material at the boundary of the craze is drawn into the craze structure at a constant σ_d, and expands by the extension ratio Λ. The middle panel of Figure 1 shows the system after the entire volume has been deformed into a craze ($L_z/L_z^0 \sim$ Λ). As the strain increases, the stress rises above σ_d until the craze fails at σ_max (bottom panel).

Compared to a homopolymer matrix, the stress-strain behavior of the composite samples shows three key differences: a reduced yield stress σ_yield in the cavitation regime (when NP-monomer interactions are purely repulsive), smaller characteristic extension ratio Λ coupled with a slightly higher drawing stress σ_d in the craze growth regime, and a maximum (failure) stress $\sigma_{max}$ in the craze breakdown regime that increases with grafted chain length N_g. In the following sections, we address these differences from a microscopic point of view.

**IV. Cavity Nucleation**

As shown in Figure 4a the polymer nanocomposites yield at smaller stretches $L_z/L_z^0$ than the homopolymer melt. The yield stress $\sigma_{yield}$ (the height of the first peak in $\sigma_{zz}$) is also significantly reduced, but is apparently independent of N_g. We emphasize that all samples in our simulations have the same thermal history; therefore, the decrease in $\sigma_{yield}$ for the composite samples is not a result of aging.

Yielding of homopolymer glasses is often associated with strain localization, or cavity nucleation.[8, 35, 36] Figure 5a compares the averaged radial polymer density around a repulsive NP before and after yielding for $M_g = 400$ (similar results are also found for $M_g = 200$ and 600). It is seen that the overall polymer density exhibits a significant depletion near the NP surface right after yielding occurs ($L_z/L_z^0 \sim 1.05$). This indicates that the repulsive interaction between NP and monomers causes cavities to nucleate preferentially near the NPs.

Further evidence of enhanced cavity nucleation near repulsive NPs is shown by plotting the radial distribution of microcavities around NPs in Figure 5b. Microcavities are defined as cubic volumes of $8\sigma^3$ containing no polymer monomers inside. The radial distribution in Figure 5b is obtained by normalizing the probability $\rho_{\text{cavity}}(r)$ of finding a cavity at radial distance r from the center of a NP by volume $4\pi r^2(2\sigma)$ at stretching factor $L_z/L_z^0 = 1.1$. There is a clear enhanced distribution of microcavities around the NP surface for the case where NP-polymer interactions are purely repulsive (i.e., employing a cutoff radius of $R_c = 11\sigma$).

One might argue that the location of nucleation merely reflects the heterogeneity induced by NPs. To test this hypothesis the interactions between the NP and the polymer were made attractive by increasing the cut-off distance $R_c$ (Eq. 4) from $11\sigma$ to $15\sigma$. As seen in Figure 5b, the cavity distribution shows a decrease near the NP when the interactions become favorable, indicating that the probability of nucleation near the NP is suppressed. This supports our conclusion that energetically unfavorable NP-polymer interfaces in composites samples reduce σ_y as compared to pure homopolymers. When the resulting yield stresses σ_y become similar in magnitude to that of pure polymer due to favorable NP-polymer interactions, cavity nucleation around NPs can be suppressed but not in the region far from NPs. In that case the resulting σ_y is expected to be comparable to that of a pure homopolymer.

For the well-dispersed composites studied here, the grafting chain length $M_g$ (and hence degree of brush-matrix and brush-brush interpenetrations) does not seem to affect the yielding behavior of composite samples. For repulsive NPs in Fig. 4a, cavitation occurs near NP-polymer interfaces, and $\sigma_{yield}$ is reduced to about the same magnitude for all values of $M_g$ studied. For attractive NP-polymer interactions shown in Fig. 4c, cavitation occurs away from NP-polymer interfaces and $\sigma_{yield}$ becomes comparable to that of pure homopolymers (this is even true for $M_g = 0$, i.e., bare NPs). These observations are consistent with the past studies on cavitation suggesting that yield stress was determined by local interactions.[11] Since NP-polymer interactions only act to change local interfacial properties, their effects are mostly manifested in the craze nucleation regime, and are not as significant in craze growth and craze break down regimes where interactions on length scales of polymer chains become relevant. Thus, the results in the following sections will be limited to composite samples with repulsive NP-polymer interactions.

## V. Craze Growth

The craze growth regime of pure homopolymers is characterized by the coexistence of crazed and uncrazed regions (Fig. 1 top), with constant corresponding polymer densities $\rho_f$ and $\rho_i$, respectively. Material in an "active zone" near the craze boundary is converted into the craze structure at the constant drawing stress $\sigma_d$.

In experiments on homopolymers, $\rho_i$ and $\rho_f$ are often measured from X-ray or electron absorption, and the ratio $\rho_i/\rho_f$ is reported as the extension ratio $\Lambda$. Figure 6a shows the average monomer density as a function of height in simulations. The low and high density regions correspond to crazed and uncrazed regions. For the pure homopolymer, $\Lambda^{\text{pure}} = \frac{\rho_i}{\rho_f} \approx \frac{0.962}{0.143} = 6.73$ which approximately matches the stretching factor beyond which the stress plateau terminates as shown in Figure 4a.

For composite samples, the different scattering of polymer and NPs complicates interpretation of x-ray data. One may expect more inhomogeneous stretching because the volume occupied by the NPs will not expand. To illustrate the nature of extension we repeat the analysis of particle displacements that was done for pure polymers in Ref. [11]. Figure 6b shows the mean final height $z_f$ of particles as a function of their initial height $z_i$. The results fall on a straight line as expected for an affine extension by a factor equal to the slope of the line. A linear fit shows that the slope is equal to $\Lambda^{\text{pure}}$ for the pure polymer system and $\Lambda^{PNC} \approx 4.95$ for both the NPs and polymer beads in the composite sample. The latter is equal to the ratio between initial and final densities of polymer beads in Fig. 6a (0.91/0.18=5.05) and to the stretching factor beyond which the stress plateau terminates in Fig. 4a. These results confirm that on large scales the composite is extended affinely by $\Lambda^{PNC} \approx 4.95$. There are small scale non-affine fluctuations due to heterogeneity. For pure polymers the fluctuations in the final height of beads from the same initial height (Fig 6c) is a half of the entanglement length. The presence of NPs increases the degree of heterogeneity and the rms variations rise about 50%. As we now discuss, the increase in heterogeneity is associated with different behavior of grafted chains.

For homopolymers, the extension ratio can be estimated by $\Lambda = \sqrt{N_e l_b l_p}$, based on scaling arguments and the assumption that chain sections become fully straightened on the length scale $N_e$ inside the crazed region. Here $N_e$ is the entanglement length, $l_b$ is the bond length and $l_p$ is the persistence length of the chains. This formula has been shown to correctly capture the qualitative trend that $\rho_f$ increases with higher entanglement density and is essentially independent of molecular weight. However, following the same logic for composite samples would imply that $N_e$ decreases by ~1.6 times. This is a surprising, unphysical outcome considering that the polymer model is unchanged – thus we expect no changes in $N_e$. Of course the NPs themselves do not expand, but since they only account for 5% of the volume, this would only reduce $\Lambda$ by about 5%.

In order to understand the decreased $\Lambda$ and increased $\rho_f$ in composite samples, we analyzed the structure inside the crazed region. The radial density of polymer segments and radial distribution of NPs around a reference NP inside the crazed region are shown in Figure 7a. The radial distributions of NPs for all three values of $N_g$ indicate a NP-NP nearest neighbor distance of ~40-50σ. Figure 6b and c shows that this reflects an increase in spacing along the extensional direction during crazing. Meanwhile the radial densities of polymers show polymer-rich regions at a distance ~15 σ from the NP surface following a polymer-depleted layer. The polymer rich regions have significantly higher polymer densities than the crazed density $\rho_f$ for a homopolymer, and reflect primarily the distortion of the brush chains. However, after excluding the polymer chains attached to a NP, Figure 7b indicates that polymer density in the region matches well with $\rho_f$ of the homopolymer and is independent of grafting chain length $N_g$. This suggests that free polymer chains as well as chain sections of grafted chains far from tethering ends (referred to as "free" sections, hereafter) behave more or less indistinguishably. However, chain sections close to tethering ends (referred to as "grafting sections", hereafter) suffer loss of degrees of freedom as a result of being connected to NPs. Upon deformation, the "grafting" sections give rise to a region of high polymer density around NPs following a depleted layer, and are responsible for the higher crazed density $\rho_f$ of the composites.

To further demonstrate the presence of the so called "grafting" sections, the bond orientation correlation function $C(\Delta N)$ is computed for grafted chains inside the crazed region.

Here, $C(\Delta N) \equiv \langle \boldsymbol{u}(\Delta N) \cdot \boldsymbol{u}(0) \rangle / \langle \boldsymbol{u}(0) \cdot \boldsymbol{u}(0) \rangle$, where $\boldsymbol{u}(0)$ is the bond vector of the tethering bond and $\Delta N$ denotes the separation from the tethering bond along chain backbone. Figure 8a shows that $C(\Delta N)$ exhibits an initially slow decay with increasing $\Delta N$. Upon further examination of the three components, it is clear that the slow decay of $C(\Delta N)$ is mostly along the extension direction z, and that appreciable decay happens only after $\Delta N \approx 20$.

Based on this observation, a series of $C(\Delta N)$ are then calculated, but with $\boldsymbol{u}(0)$ being the bond vector of the $\delta N$th bond from the tethering site as shown in Figure 8b. For $\delta N = 100$, $C(\Delta N)$ essentially overlaps with that of free chains in the sample, exhibiting significantly faster decay compared to $C(\Delta N)$ for $\delta N = 0$. This reinforces our conjecture that "grafting" sections on the brush chains have very different conformations than the rest of the grafts. The length scale of the "grafting" sections can be roughly estimated from the number $\delta N$ required to recover the $C(\Delta N)$ of ungrafted chains, approximately ~100 bonds in this case.

In addition to bond correlations, the differences in microstructure of "grafting" sections are also reflected by the scaling of the z component of the mean square distance between two segments separated by $\Delta N$ bonds $\langle z^2(\Delta N) \rangle$ with respect to $\Delta N$. As shown for $N_g = 400$ in Figure 8c, $\langle z^2(\Delta N) \rangle \propto \Delta N^2$ starting from the tethering points up to $\Delta N \sim 15$. This suggests that the first ~15 segments starting from tethering sites are pulled taut along the z direction. This length scale is consistent with the distance from NP surfaces over which polymer-rich regions are located (Figure 6a). Once again, the scaling behavior of "free" sections on grafted chains are similar to the free chains, both of which are much less stretched than the "grafting" sections.

The FENE bond tensions along backbones of grafted and free chains inside the crazed region are shown in Figure 8d for the sample with $N_g = 400$. The tension along chain backbone is calculated by recording the component of the stress tensor in the pulling direction for each segment due to FENE bond during extension, followed by an average over segments of the same location along chain contour. Only segments in the crazed region are included. For segments far from the tethering ends of the grafted chains (i.e., "free" sections), the tensions are nearly identical to that of free chains, exhibiting a plateau in the interiors of chains and decaying toward zero as free ends are approached. In fact, tensions also match quantitatively with what is measured for the homopolymer and are due to "inter-connectivity" of those chain sections through entanglements. The presence of "grafting" sections on grafted chains is now clearly manifested by the elevated tensions when tethering ends are approached. The length scale of "grafting" sections is consistent with what is suggested from bond correlations, i.e. ~100 segments. The tension analysis proves that during craze growth "free" sections on grafted chains respond to deformation in a similar way as free chains, and chains in the corresponding homopolymer. It is the presence of "grafting" sections that are responsible for most of the differences between tethered NP-polymer composites and homopolymer. In the case of tensions, this is reflected by a higher drawing stress $\sigma_{draw}$ for the composites samples as shown in Figure 4(a).

Figure 8a-d unambiguously show that the loss of degrees of freedom in motion of "grafting" sections of ~100 monomers plays a prominent role in changing the mechanical response of the grafted-NP composites samples as compared to the homopolymer. As a result, the removal of the grafting constraints should yield mechanical behavior akin to the homopolymer, which is indeed the case as indicated by Figure 4c.

## VI. Craze Failure

Stretching samples beyond $L_z/L_z^0 = \Lambda$ requires deformation of the entanglement network and results in stress increases that mostly arise from the stretching of covalent bonds. Eventually, this leads to catastrophic failure either through bond breaking or chain disentanglement (Fig. 1 bottom). These two competing mechanisms determine the maximum stress $\sigma_{max}$ that the samples can sustain before failure. The fracture energy for crack propagation scales as the square of σ_max.[10, 37] In general, $\sigma_{max}$ is a function of polymer chain length but saturates for chain lengths greater than 5-10Ne.[38] Simulations[11, 12] show that the saturation of stress is associated with a transition from chain disentanglement at small N to chain scission at large N. For shear failure of polymer glasses, the transition from chain pullout to chain scission occurs at almost the same chain length in simulations[39] of polymers with different $N_e$, suggesting that chain friction is critical in determining the chain length in which chain pullout dominates over chain scission.[40]

For tethered NP-polymer composites, as discussed in Section V, the "grafting sections" bear significantly higher bond tensions than other polymer segments. Figure 9a shows that this is still true beyond the growth regime ($L_z/L_z^0 > \Lambda$), except for $N_g = 200$. Comparing to the craze growth regime, as $L_z/L_z^0 > \Lambda$, bond tensions for both tethered and matrix chains are observed to increase with stretching factor, while tensions near free ends still vanish as a result of "free-end disentanglements" from tube escape. The chain contour length that is subject to "free-end disentanglements" can be estimated by fitting the measured bond tension to a function $a_1[1 - exp(-N/a_0)]$ (for $N \lesssim 10$). The fitting parameter $a_0$ increases from $\sim N_e$ to $\sim 5N_e$ as stretching factor $L_z/L_z^0$ increases from 4.95 to 9.5, suggesting an increasing degree of free-end disentanglement as samples are stretched beyond $\Lambda$. It is imaginable that for samples with $N_g \lesssim 5N_e$ (e.g. $N_g = 200$) bond tensions on the "grafting sections" will eventually be reduced as the range of free-end disentanglement progresses toward the tethering ends, as shown in Figure 9a. However, for samples with $N_g \gtrsim 5N_e$, (e.g. $N_g = 400$ and 600) this effect becomes much less obvious. For free polymer chains, disentanglements occur at both ends, resulting in lower "total" bond tensions than that on grafted chains. It is worth noting that tensions on free chains are found to be nearly identical for all three composite samples ($N_g$ = 200, 400 and 600, data not shown), indicating that the asymptotic $N_g$-independent limit is reached.

Intuitively, in the craze breakdown regime one may expect entanglements formed between brushes belonging to different NPs to play a defining role in determining the observed $N_g$ dependent stress-strain behavior (Figure 4a). However, our data do not support this idea for the following two reasons. First, given the different degree of brush-brush interpretations of the three composites samples (Figure 3a-c), Figure 9a shows that free-end disentanglements are found to vary with $L_z/L_z^0$ in a very similar way. This suggests that entanglements by brush-brush and brush-matrix interpenetrations are not distinguishable in regard to releasing free-ends. Second, as long as $N_g$ is greater than the contour range of free-end disentanglements, similar tensions are obtained on the "grafting sections" ($N_g = 400$ and 600 in Figure 9a), further indicating that neither the degrees of brush-brush and brush-matrix entanglements affect tensions on "grafting sections". Instead, according to the analysis in the previous paragraph, higher tensions are carried by grafted chains due to the losses of half of the free ends. In the craze failure regime, the faster increase in stress for samples with larger $N_g$ can be simply understood as a result of higher volume fractions of grafted polymers.

The dominant failure mechanism can be inferred from Figure 9a. As bond tensions increase with increasing range of free-end disentanglements, an upper limit in the range is expected. Beyond this, bond tensions exceed the bond breaking threshold ($100 \times$ LJ bonds in this study). When the grafting chain length $N_g$ is short compared to this upper limit, the system fails predominantly through disentanglements (e.g. $N_g = 200$), whereas for large $N_g$ the sample fails via bond breaking ($N_g = 400$ and 600). To confirm this, the fractions of broken bonds in different samples are shown in Figure 9b as a function of stretching factor. All data can be well fitted using a hyperbolic-tangent function of the form $A \times tanh(x)$ with $x = c(L_z/L_z^0 - \lambda_{max})$. The asymptotic fraction of broken bonds $A = 4.1 \times 10^{-5}$, $2.3 \times 10^{-4}$ and $4.1 \times 10^{-4}$ for $N_g = 200, 400$ and 600, respectively. Moreover, for $N_g = 600$ and 400 but not 200, the stretching factor with the highest rate of bonds breaking (per stretching) also produces $\sigma_{max}$ in the stress-strain curves shown in Figure 4a. An analysis of the distribution of the broken bonds indicates that >95% of the broken bonds belong to grafting chains (data not shown). Similar to what has been reported for homopolymers,[12] catastrophic failure requires only a tiny fraction of bonds being broken (0.04% for $N_g = 600$). The state of "disentanglement" and "bond breaking" can also be indicated by snapshots taken at $L_z/L_z^0 = 14.0$ as shown in Figure 9(c). The snapshots show that grafted chains from two NPs are "parting away" without breaking for simulations with $N_g = 200$, while extensive entanglements are still present for simulations with $N_g = 600$. Note that bond breaking is not directly visible due to small number of broken bonds.

One important experimentally measurable quantity is the macroscopic fracture energy $G_c$. During tensile fracture of many polymers, large volumes of material around the advancing crack are deformed into a craze.[7,8] The work required to deform this material greatly enhances $G_c$. Direct simulations of this process are not possible, since the width of crazed regions is micrometers and the crack length is millimeters.[7,8] However, Rottler et al.[10] showed that fracture energies for homopolymers could be obtained by using small scale simulations to determine the parameters of a macroscopic fracture model.[37] In particular,

$$G_c = 4\pi\kappa D_0 \frac{\sigma_{max}^2}{\sigma_{draw}}(1 - 1/\Lambda)$$

where $D_0$ is the fibril spacing and the prefactor $\kappa$ depends on the anisotropic elastic constants of the crazed network but is typically 1 to 3. This expression allows us to determine what factors will optimize the fracture energy of composites. As both $\sigma_{draw}$ and $1/\Lambda$ are greater for samples filled with polymer grafted NPs, higher fracture energy can only be achieved via increasing $\sigma_{max}$. According to our study, this requires the grafting chain length $N$ being larger than the maximum range of "free-ends disentanglements" that itself depends on $N_e$ and covalent bond strength.

## VI. Conclusions

In this work, molecular dynamics simulations were used to study the crazing behavior of polymer nanocomposites comprised of polymer-grafted nanoparticles and homopolymers. The polymer chain lengths studied are well into the entanglement regime. Comparing to a previous study on crazing of pure homopolymers using a similar model,[10, 11] crazing composite samples exhibit differences in the craze nucleation, growth, and craze break down regimes. We show that the yielding stresses $\sigma_y$ of polymer nanocomposites are controlled by the NP-polymer interfacial properties. Unfavorable NP-polymer interactions reduce $\sigma_y$, while favorable interactions restore $\sigma_y$ to that of the pure homopolymer. The craze growth regime of composite samples is characterized by inhomogeneous polymer segmental densities around NPs in crazed regions. Comparing to pure homopolymer samples, polymer segmental density is depleted around NPs, followed immediately by an enriched layer which is the result of the loss of degrees of freedom of the "grafting sections". This results in a reduced extension ratio $\Lambda$ as compared to pure homopolymers. A bond tension analysis also shows that in the crazed region the "grafting sections" carry significantly higher tension than free polymers, which is responsible for the higher drawing stress $\sigma_d$ exhibited by the composite samples.

As in pure homopolymers, there are two competing failure mechanisms, disentanglements and bond breaking, respectively. In the craze failure regime, with increasing stretching factor the contour length of disentangled chain sections increases as do bond tensions along the polymer backbone. Therefore, there exists a "lower limit" in chain length beyond which failure via disentanglements will be preempted by bond breaking. The previous study shows that this limit is ~$20N_e$ in pure homopolymers samples[39] above which the failure stress $\sigma_{max}$ becomes chain length independent. For composites, since the "grafting sections" carry much higher bond tensions than free chains, the "lower limit" is set by the grafting chain length. Our results suggest that bond breaking on grafting chains occurs as "free-end disentanglements" reach $\sim 5N_e$. As a result, composite samples with grating chain length $N_g \gtrsim 5N_e$ fail predominantly through bond breaking, as found for $N_g = 400$ and 600 but not 200. Saturation of failure stress $\sigma_{max}$ with respect to $M_g$ is observed for all the samples studied. This is probably because of the increase in the number of segments on grafting chains that carry higher bond tensions.

**Acknowledgements**
Financial support from the National Science Foundation (CBET-1033168) and (DMR-1411144) is gratefully acknowledged. This research used resources of the National Energy Research Scientific Computing Center, which is supported by the Office of Science of the U.S. Department of Energy under Contract No. DE-AC02-05CH11231 and the Advanced Scientific Computing Research (ASCR) Leadership Computing Challenge (ALCC). This work was performed, in part, at the Center for Integrated Nanotechnology, a U.S. Department of Energy and Office of Basic Energy Sciences user facility. Sandia National Laboratories is a multi-program laboratory managed and operated by Sandia Corporation, a wholly owned subsidiary of Lockheed Martin Corporation, for the U.S. Department of Energy's National Nuclear Security Administration under contract DE-AC04-94AL85000.

**Figures and Captions**

Figure 1: Snapshots from simulations showing (a) craze growth, (b) fully formed crazed, and (c) failure for homopolymer (left, with $L_z/L_z^0 = 4.0, 9.0,$ and $14.5,$ respectively) and tethered NP-polymer composite (right, with $L_z/L_z^0 = 4.0, 14.0,$ and $19.0,$ respectively).

Figure 2: Density ρ as function of temperature during cooling at constant pressure P=0. The glass transition temperature $T_g$ is determined from the crossing of the linear extrapolation from high and low temperatures;

Figure 3: Radial densities of polymer segments around tethered NPs with (a) $N_g$=200, (b) 400 and (c) 600 with polymer/NP interactions cutoff at $R_c = 11\sigma$: overall polymers (black), own brushes (red), grafted chains on other NPs (green) and free chains (blue). The black dotted lines indicate the bulk density of homopolymer at T=0.2.

Figure 4: Stress-strain curves of (a) homopolymer and tethered NP-polymer composites with purely repulsive polymer/NP interactions ($R_c = 11\sigma$); (b) homopolymer, tethered NP-polymer composite ($N_g = 200$) with purely repulsive ($R_c = 11\sigma$) and attractive ($R_c=15\sigma$) NP-polymer interactions; (c) Homopolymer, tethered NP-polymer composites ($N_g$=200) and bare NP composites with attractive ($R_c = 15\sigma$, $A_{cs} = 80$) NP-polymer interactions.

Figure 5: (a) Radial densities of overall polymer segments $\rho_{poly}$ (symbol line) and brush segments $\rho_{brush}$ (dashed line) before ($L_z/L_z^0 = 1.02$ in red) and after ($L_z/L_z^0 = 1.1$ in blue) yielding; (b) Probability of finding a cavity cell (see text for definition) as the function of distance from NPs at $L_z/L_z^0=1.1$.

Figure 6: (a) Polymer density profiles along the stretching direction of homopolymer (blue circles) and tethered NP-polymer composites (green squares) ($N_g$=400); (b) final height $z_f$ as a function of initial height $z_i$, and (c) standard deviation from the average height in each layer, of segments in homopolymer (blue circles), segments in tethered NP-polymer composites (green squares), and NPs in polymer composites (red triangles).

Figure 7: (a) Radial densities for polymer around a reference tethered NP inside the crazed region (open circles for total polymer segments, dashed lines for segments on own brushes, and open triangles for all other segments); filled circles shows NP-NP radial distribution function $g(r)$. Red, green, blue for composites sample with $N_g$ = 200, 400, 600, respectively. The black dashed line indicates the crazed density for homopolymer. (b) Similar data with the grafted chains removed.

Figure 8: (a) Bond correlation function and its components calculated for the grafted chains inside crazed region ($N_g$ = 400); (b) bond correlation functions calculated for free chains and grafted chains inside crazed region with $\delta N$ = 0 and 100; (c) scaling of $\langle \Delta z^2 \rangle / \Delta N^2$ with respect to $\Delta N$ (see text for definition) for grafted chains and free chains inside crazed region; (c) distribution of FENE bond tensions along chain backbone for polymers inside crazed region; grafted chains (red), free chains (blue) and homopolymer (green).

Figure 9: (a) Comparison of FENE bond tensions along chain backbones of grafted chains ($N_g$=600 (blue), 400 (green), 200(red)) and free chains (pink), at $L_z/L_z^0 \sim$ 4.95 (crosses) and 9.5 (solid circles). (b) Fraction of broken bonds as a function of extension ratio $L_z/L_z^0$ ($N_g$=600 in blue, 400 in green, 200 in red and homopolymer in black). Dotted lines show the fitted functions. (c) Snap shots from simulations showing configurations of grafted chains on two neighboring NPs (distinguished in blue and red color), at extension ratio $L_z/L_z^0$=12.0. The upper and lower plots are from simulations with $N_g$=200 and 600, respectively.

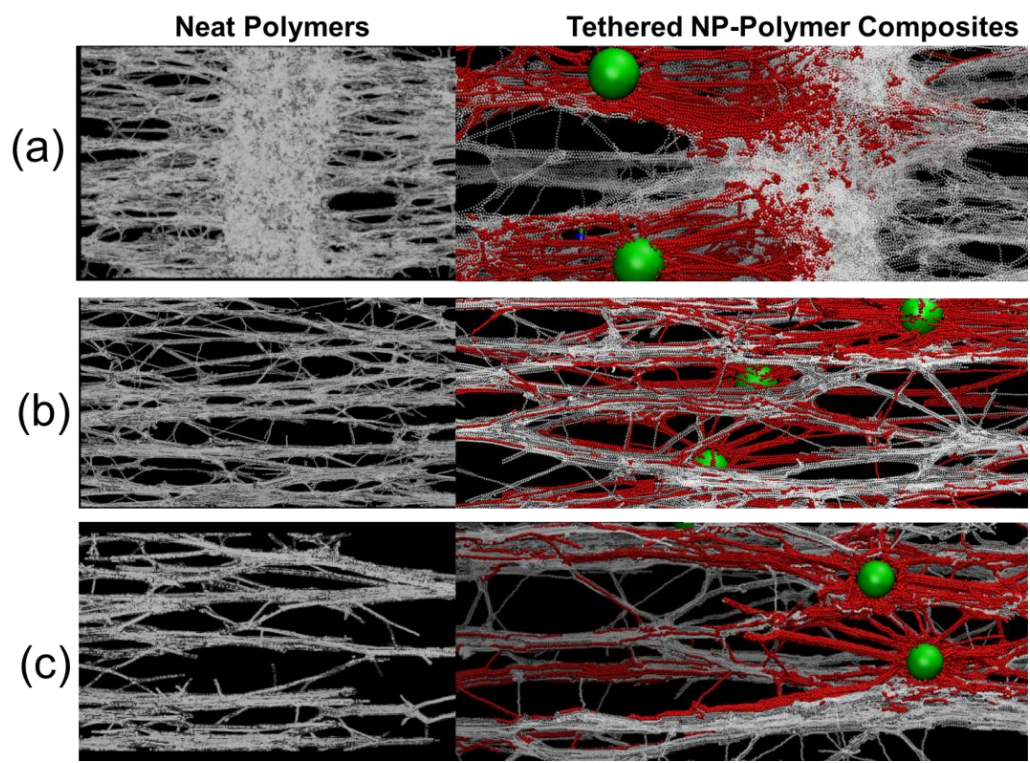

**Figure 1**

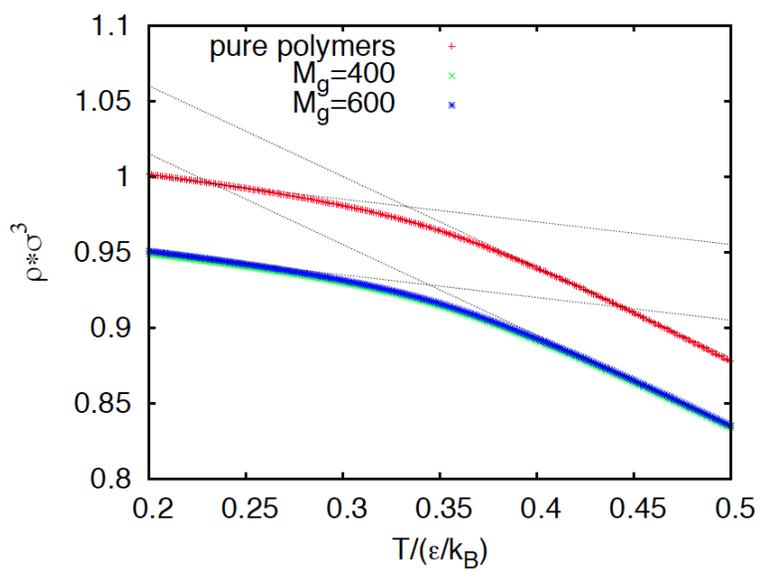

**Figure 2**

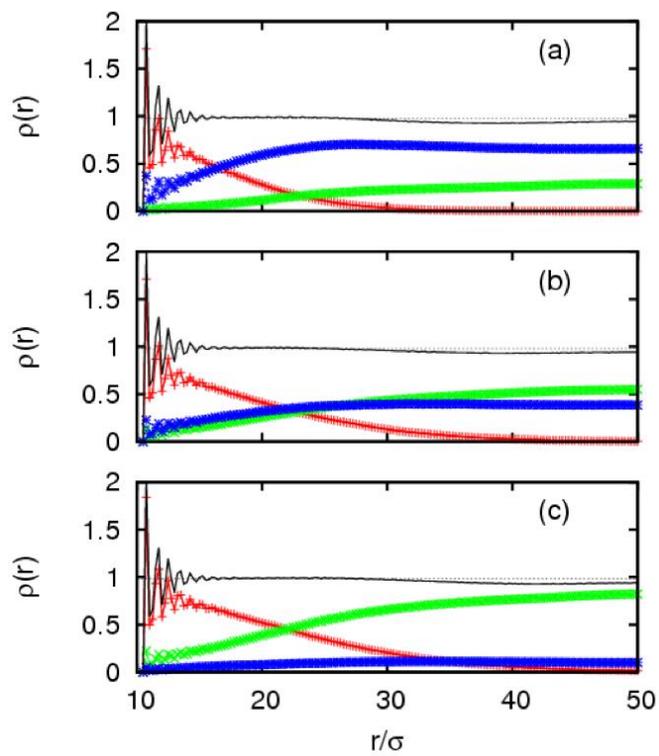

**Figure 3 (a), (b) and (c)**

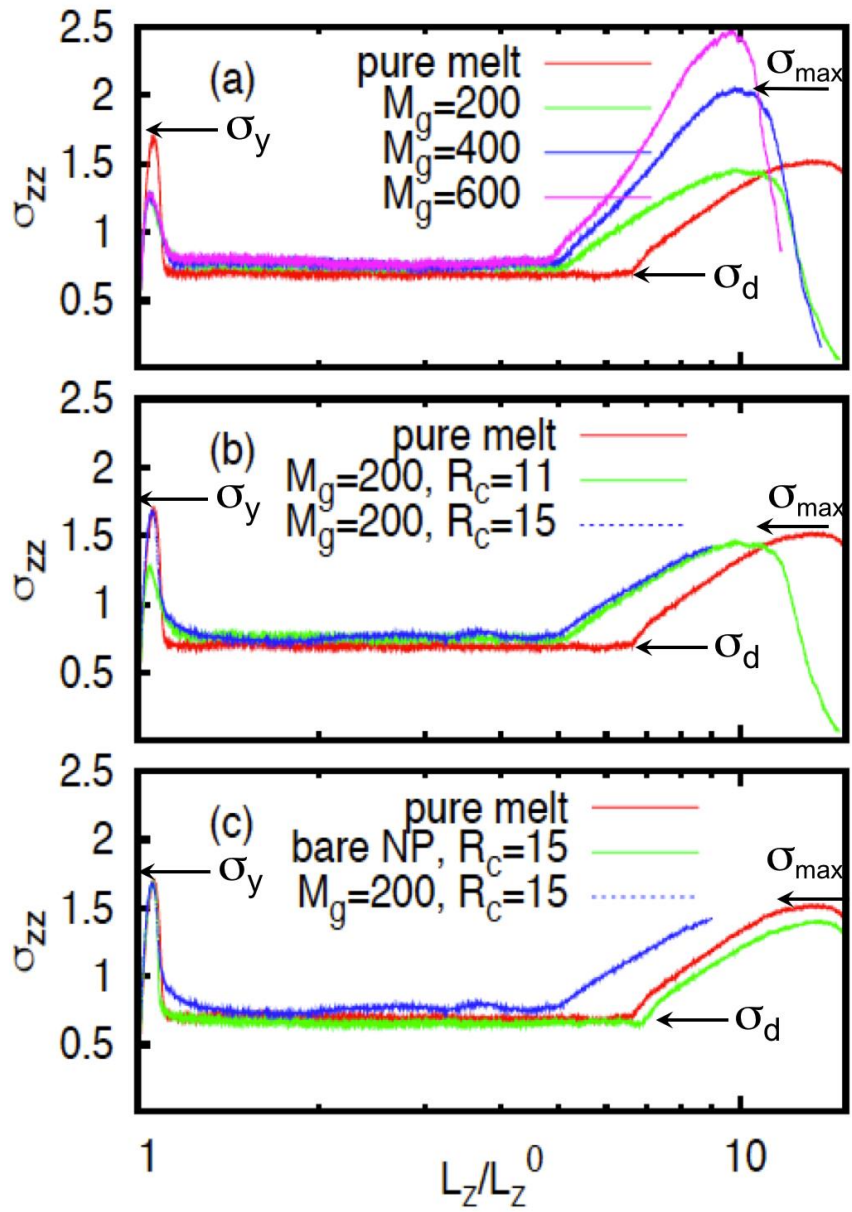

**Figure 4** (a), (b), (c)

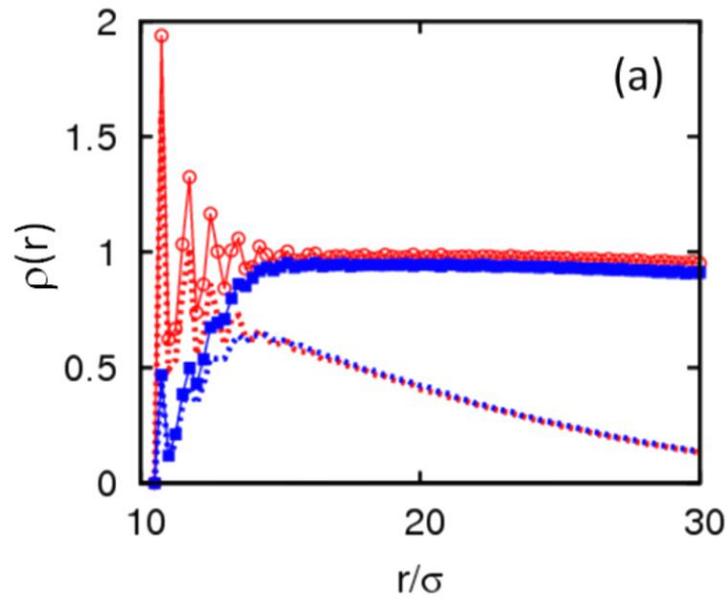
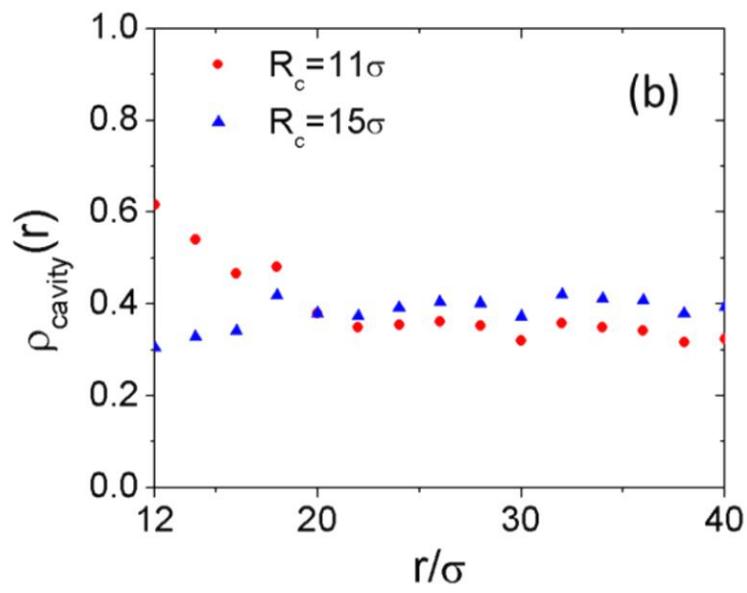

**Figure 5 (a) and (b)**

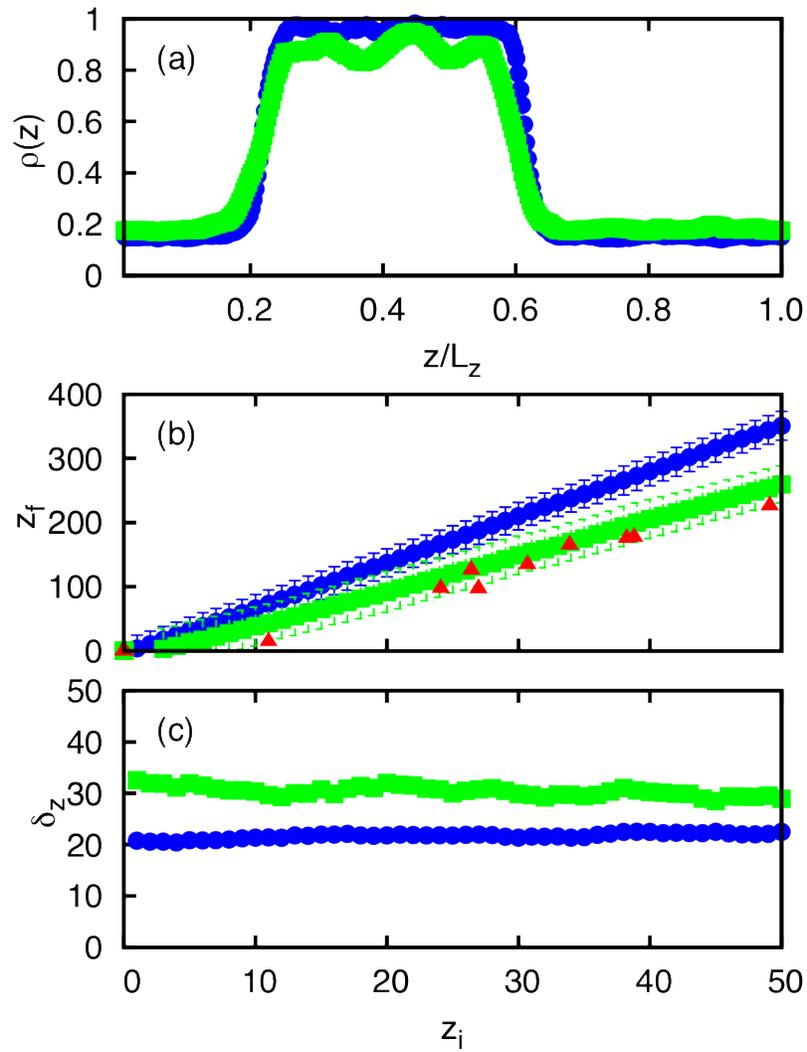

**Figure 6 (a), (b) and (c)**

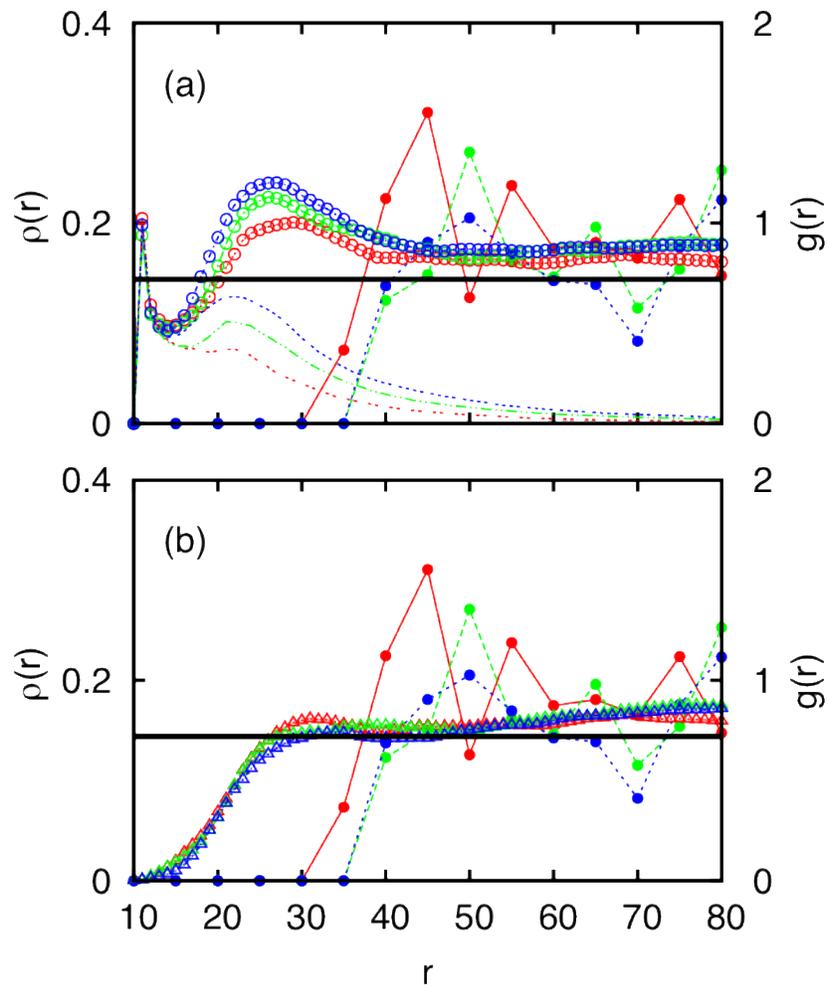

**Figure 7 (a) and (b)**

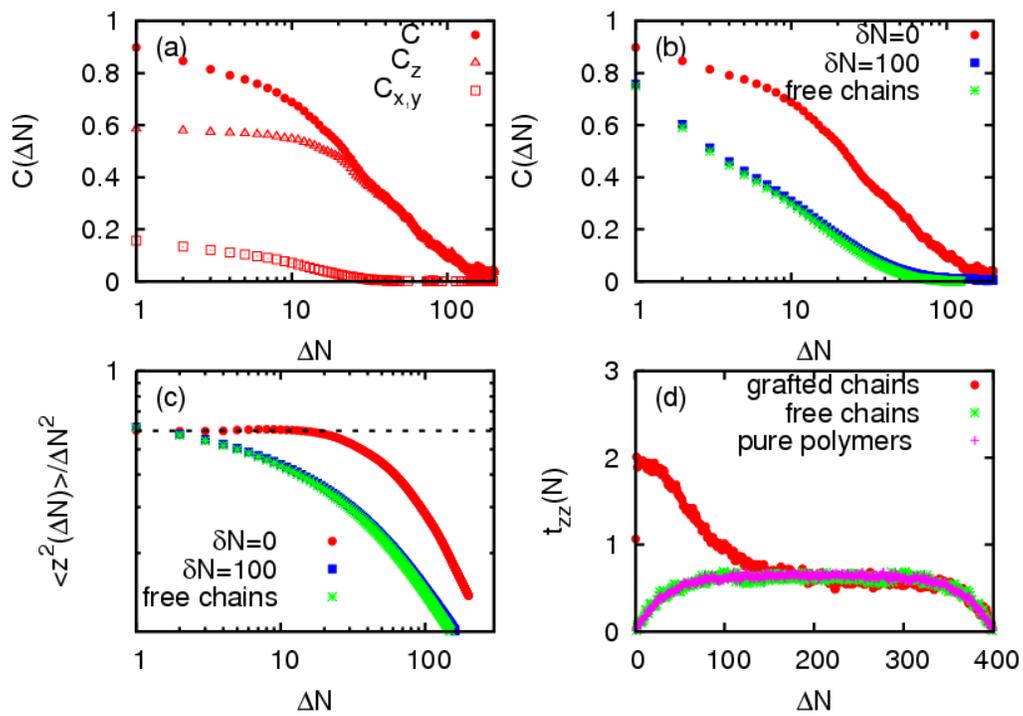

**Figure 8 (a), (b), (c) and (d)**

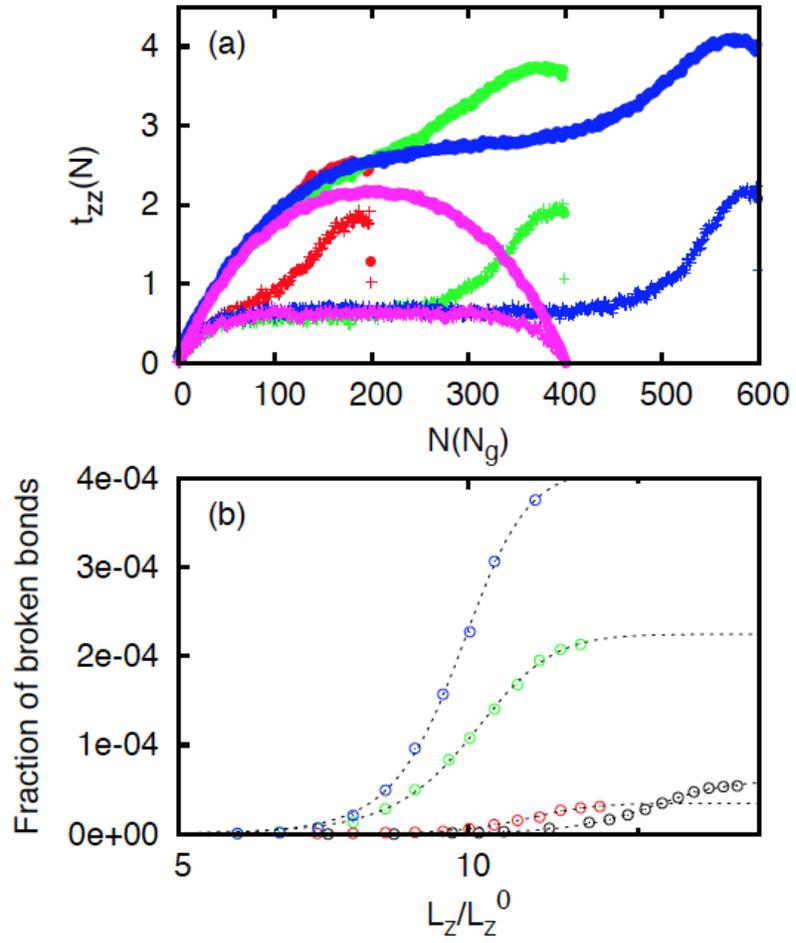

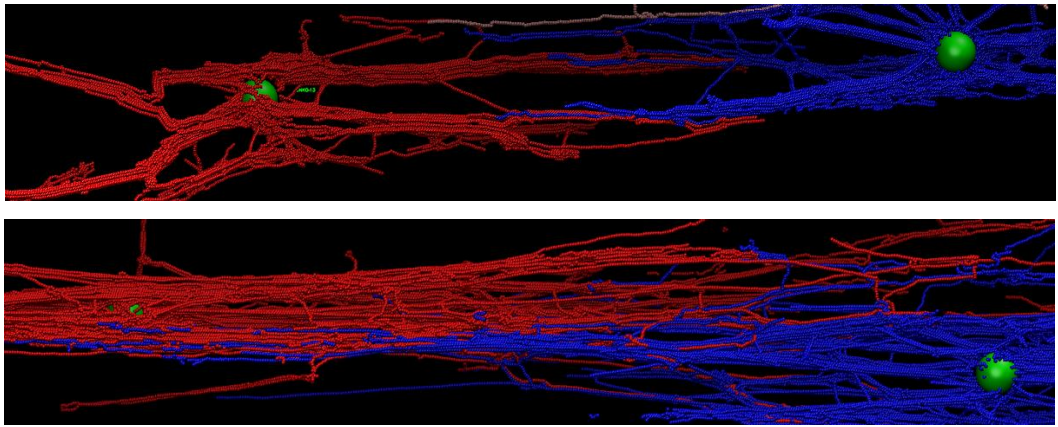

**Figure 9 (a), (b) and (c)**